\documentclass[aps,prl,twocolumn,showpacs, superscriptaddress]{revtex4-1}
\usepackage{graphicx}
\usepackage{latexsym}
\usepackage{amsmath}
\usepackage{lmodern}

\begin{document}

\title{Doping Evolution of Magnetic Order and Magnetic Excitations in (Sr$_{1-x}$La$_x$)$_3$Ir$_2$O$_7$}

\author{Xingye Lu}
\email{xingye.lu@psi.ch}

\author{D. E. McNally}
\affiliation{Research Department Synchrotron Radiation and Nanotechnology, Paul Scherrer Institut, CH-5232 Villigen PSI, Switzerland}

\author{M. Moretti Sala}
\affiliation{European Synchrotron Radiation Facility, BP 220, F-38043 Grenoble Cedex, France}

\author{J. Terzic}
\affiliation{Department of Physics and Astronomy, University of Kentucky, Lexington, Kentucky 40506, USA}
\affiliation{Department of Physics, University of Colorado at Boulder, Boulder, CO 80309}
\author{M. H. Upton}
\author{D. Casa}
\affiliation{Advanced Photon Source, Argonne National Laboratory, Argonne, Illinois 60439, USA}

\author{G. Ingold}
\affiliation{Research Department Synchrotron Radiation and Nanotechnology, Paul Scherrer Institut, CH-5232 Villigen PSI, Switzerland}
\affiliation{SwissFEL, Paul Scherrer Institut, CH-5232 Villigen PSI, Switzerland}

\author{G. Cao}
\affiliation{Department of Physics and Astronomy, University of Kentucky, Lexington, Kentucky 40506, USA}
\affiliation{Department of Physics, University of Colorado at Boulder, Boulder, CO 80309}

\author{T. Schmitt}
\email{thorsten.schmitt@psi.ch}
\affiliation{Research Department Synchrotron Radiation and Nanotechnology, Paul Scherrer Institut, CH-5232 Villigen PSI, Switzerland}

\date{\today}

\begin{abstract}
We use resonant elastic and inelastic X-ray scattering at the Ir-$L_3$ edge to study the doping-dependent magnetic order, magnetic excitations and spin-orbit excitons in the electron-doped bilayer iridate (Sr$_{1-x}$La$_{x}$)$_3$Ir$_2$O$_7$ ($0 \leq x \leq 0.065$). With increasing doping $x$, the three-dimensional long range antiferromagnetic order is gradually suppressed and evolves into a three-dimensional short range order from $x = 0$ to $0.05$, followed by a transition to two-dimensional short range order between $x = 0.05$ and $0.065$. Following the evolution of the antiferromagnetic order, the magnetic excitations undergo damping, anisotropic softening and gap collapse, accompanied by weakly doping-dependent spin-orbit excitons. Therefore, we conclude that electron doping suppresses the magnetic anisotropy and interlayer couplings and drives (Sr$_{1-x}$La$_x$)$_3$Ir$_2$O$_7$ into a correlated metallic state hosting two-dimensional short range antiferromagnetic order and strong antiferromagnetic fluctuations of $J_{\text{eff}} = \frac{1}{2}$ moments, with the magnon gap strongly suppressed.
\end{abstract}


\maketitle

The layered Ruddlesden-Popper iridate series Sr$_{n+1}$Ir$_n$O$_{3n+1}$ ($n = 1, 2$) that hosts novel $J_{\text{eff}}=\frac{1}{2}$ Mott insulating states have recently attracted much interest owing to their potential for exploring novel collective quantum states by charge-carrier doping in the strong spin-orbit coupling (SOC) limit \cite{jackeli_prl, bjkim_08, bjkim_09, dimensionality_mit, jwkim_327, balents10, review14, review15}. Distinct from $3d$ Mott insulators where strong on-site Coulomb electron correlation ($U$) dominates \cite{rmp_06}, the $J_{\text{eff}}=\frac{1}{2}$ Mott state in iridates is induced by cooperative interplay between crystal-field, SOC ($\sim 0.4$ eV) and an intermediate $U$ \cite{bjkim_08}. The novel Mott insulator Sr$_2$IrO$_4$ is similar to La$_2$CuO$_4$ in magnetic order, spin dynamics and electronic structure \cite{bjkim_08,jackeli_prl,bjkim_09,fye_214,dimensionality_mit,review15,jkim_214, bjkim_14ncom, radu_01,headings_10}. It has been suggested that electron-doped Sr$_2$IrO$_4$ is an analogous system to hole-doped La$_2$CuO$_4$ \cite{wangfa_2011, mc_2013}, which is supported by numerous nontrivial experimental observations such as Fermi arcs, pseudogaps, $d$-wave gaps and persistent paramagnons \cite{cao_la214, wilson_214, 214_fermiarc, donglai, 214_dwave, La214_arpes, hlynur_la214, xliu_la214}.

Compared with Sr$_2$IrO$_4$, the bilayer Sr$_3$Ir$_2$O$_7$ bearing strong interlayer coupling is similar to bilayer cuprates \cite{ybco} and shows unique advantages in exploring novel phases via electron doping since it retains the $J_{\text{eff}}=\frac{1}{2}$ Mott state while lying close to an insulator-to-metal transition (IMT) \cite{dimensionality_mit,jwkim_327}. Due to the bilayer structure, Sr$_3$Ir$_2$O$_7$ exhibits a much smaller gap $\Delta E \approx 130$ meV than Sr$_2$IrO$_4$ ($\Delta E \approx 600$ meV) \cite{214_stm, 327_stm, dimensionality_mit}. The strong interlayer couplings and magnetic anisotropy including the bond-directional pseudodipolar interactions arising from the $J_{\text{eff}}=\frac{1}{2}$ states induce a $c$-axis $G$-type antiferromagnetic order (AFM) below $T_{\text{N}} \approx 285$ K \cite{jwkim_327, stefano_12} and distinct magnons bearing a large magnon gap $\approx 90$ meV \cite{jkim_327, marco_327}. Because of the small charge gap of Sr$_3$Ir$_2$O$_7$, one can expect a homogeneous metallic state to develop by minor charge carrier doping. Indeed, an IMT and a robust metallic state have been realized in (Sr$_{1-x}$La$_x$)$_3$Ir$_2$O$_7$ for $x\gtrsim 0.05$ \cite{cao_la327,wilson_327}. Since minor La dopants have little structural effect on IrO$_2$ layers while driving the system across the IMT, (Sr$_{1-x}$La$_x$)$_3$Ir$_2$O$_7$ provides an ideal platform to exploring novel phenomena arising from charge carrier doped $J_{\text{eff}} = \frac{1}{2}$ Mott states, in the presence of strong interlayer couplings and pseudodipolar interactions \cite{jkim_327, marco_327}. However, the doping evolution of the ground state of (Sr$_{1-x}$La$_x$)$_3$Ir$_2$O$_7$ is still under intense debate \cite{wilson_327,La327_arpes,junfeng_sr,junfeng_nmat}. Angle-resolved photoemission (ARPES) measurements revealed a strong coherent quasiparticle peak and suggested a weakly correlated Fermi liquid ground state in (Sr$_{0.95}$La$_{0.05}$)$_3$Ir$_2$O$_7$ \cite{La327_arpes}. In contrast, a doping-dependent negative electronic compressibility was discovered in later ARPES measurements, indicating an exotic correlated metallic state \cite{junfeng_nmat, wilson_327}. In order to reveal the nature of the metallic state, a detailed study of the elementary excitations is required that can determine the doping evolution of the electronic interactions, especially the magnetic couplings. Furthermore, it will be interesting to study the doping dependence of the strong interlayer couplings and the pseudodipolar interactions that drive the AFM \cite{jwkim_327} and the large magnon gap in Sr$_3$Ir$_2$O$_7$ \cite{jkim_327,marco_327}.

In this letter, we present measurements of the doping dependence of the magnetic order and the elementary excitations across the electron-doping driven IMT in (Sr$_{1-x}$La$_x$)$_3$Ir$_2$O$_7$ ($x = 0, 0.02, 0.03, 0.05$ and $0.065$) using Ir $L_3$ edge resonant inelastic X-ray scattering (RIXS) \cite{rmp_rixs}. Our results reveal an evolution of the AFM from three-dimensional (3D) long range AFM (LAF) ($x \leq 0.03$) to 3D short range AFM (SAF) ($x \sim 0.05$) across the IMT and subsequent 2D SAF deep in the metallic state ($x = 0.065$). Following the evolution of the magnetic order, we present a detailed analysis of the doping-dependent magnetic excitations and spin-orbit excitons, from which the doping effects on the magnetic couplings and the nature of the metallic state are determined.

\begin{figure}[htp]
\includegraphics[width=8.7cm]{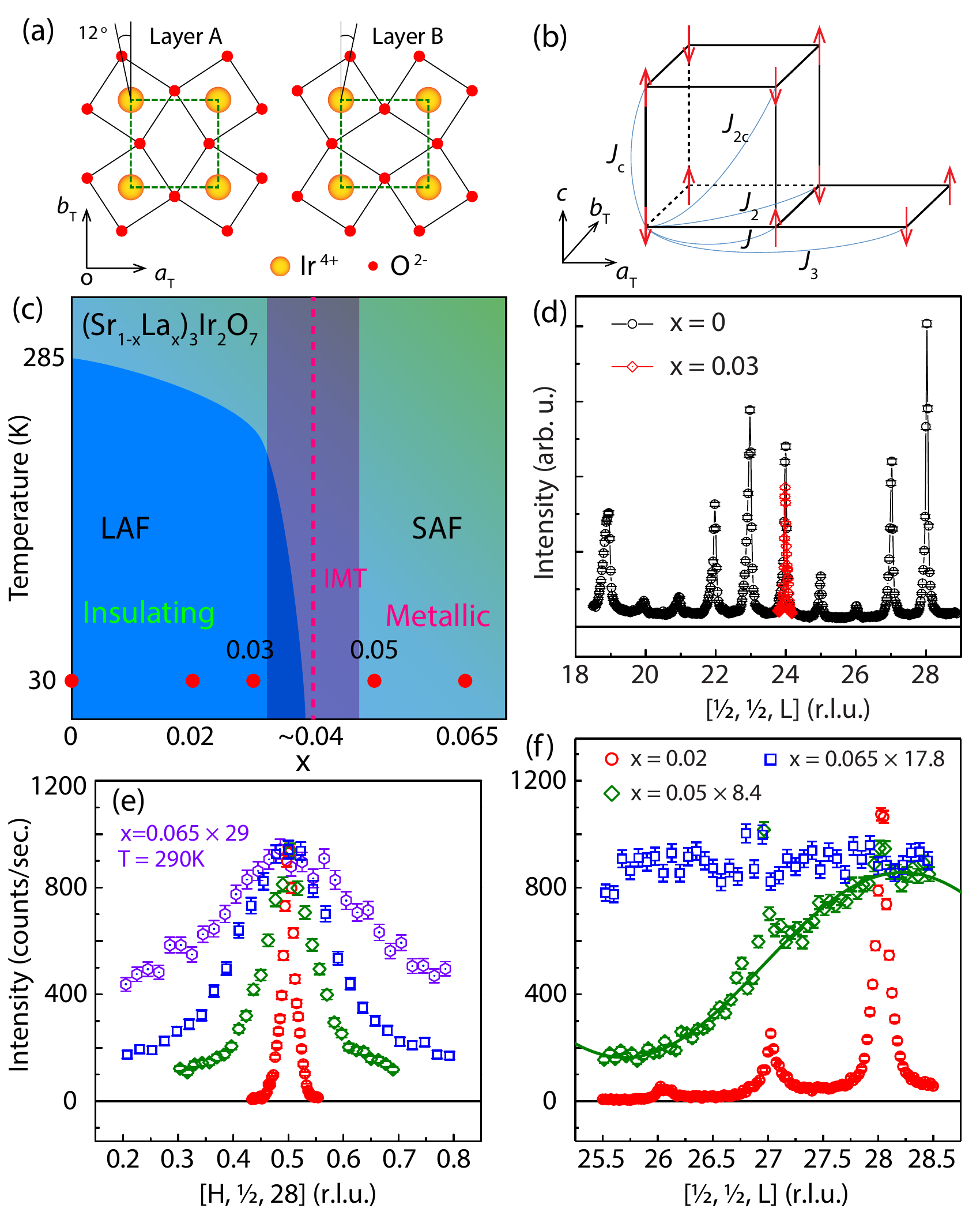}
\caption{(color online). (a) In-plane crystal structure and the rotation of the IrO$_6$ octahadra ($\sim 12^{\circ}$) around $c$ axis. (b) $G$-type collinear antiferromagnetic order with moment along $c$ axis . $J$, $J_2$ and $J_3$ are first, second and third nearest superexchange couplings within $ab$ plane, respectively. $J_c$ and $J_{2c}$ are first and second nearest couplings along $c$ axis. (c) Schematic phase diagram of (Sr$_{1-x}$La$_x$)$_3$Ir$_2$O$_7$, adapted from ref. \cite{cao_la327,wilson_327}. LAF and SAF are long range and short range antiferromagnetic order, respectively. IMT is insulator-to-metal transition, which occurs at $x \sim 0.04$. The dopings used in the present study are marked by red dots. (d) $L$ scan of the $c$-axis $G$-type antiferromagnetic order for $x = 0$ and $0.03$. (e),(f) Doping dependent $H$ and $L$ scans across the magnetic Bragg peak ($\frac{1}{2}, \frac{1}{2}, 28$). The green solid curve in (e) is a fit of the $L$ scan for $x = 0.05$. All the measurements were performed at 30 K unless otherwise indicated.}
\label{fig:1}
\end{figure}

\begin{figure*}[htbp]
\includegraphics[width=16cm]{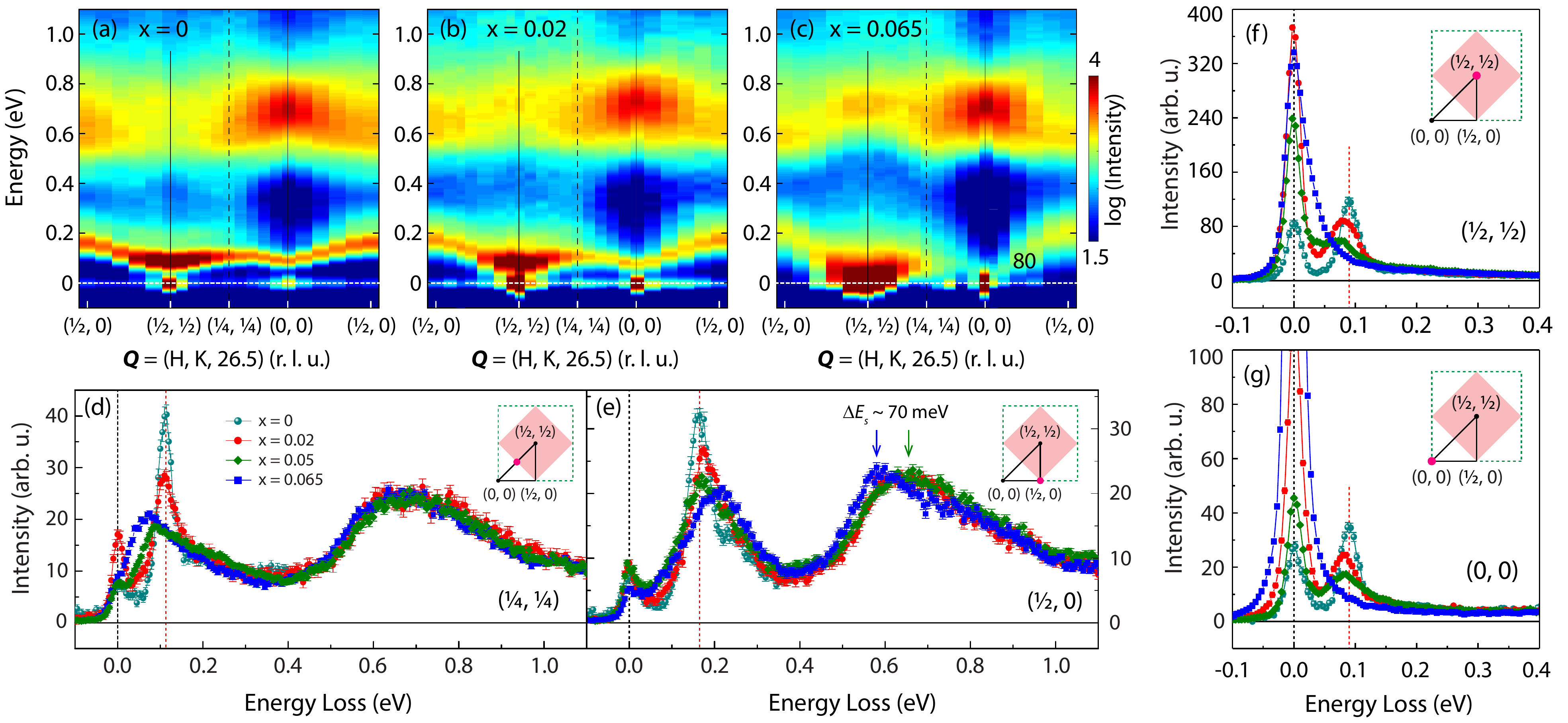}
\caption{(color online). (a)-(c) In-plane momentum dependence of RIXS spectra of (Sr$_{1-x}$La$_x$)$_3$Ir$_2$O$_7$ for $x=0$, $x=0.02$ and $0.065$. (d)-(e) Comparison of the elementary excitations between $x=0, 0.02, 0.05$ and $0.065$ collected at $\textbf{Q} = (\frac{1}{4}, \frac{1}{4})$ and $(\frac{1}{2}, 0)$. For $x = 0$, only magnetic excitations are shown. (f)-(g) Doping-dependent magnetic excitations at $\textbf{Q} = (\frac{1}{2}, \frac{1}{2})$ and (0, 0). The insets in (d)-(g) illustrate the reciprocal space where the green dashed square and the pink solid square are the tetragonal Brillouin zone and the AFM Brillouin zone, respectively. The pink filled circles mark the vector positions for (d)-(g). The vertical red dashed lines mark the peak position of the magnons for $x=0$. The blue and green arrows in (e) show the peak positions of the spin-orbit excitons and $\Delta E_s$ marks the energy difference of the peak positions.}
\label{fig:2}
\end{figure*}

 The measurements on (Sr$_{1-x}$La$_x$)$_3$Ir$_2$O$_7$ were carried out at the ID20 beamline ($x = 0.02, 0.05$ and $0.065$) of the European Synchrotron Radiation Facility (ESRF) and the 27ID-B beamline ($x = 0$ and $0.03$) at the Advanced Photon Source (APS) \cite{marco_id20,27idb, SI}. To facilitate comparison with previous results \cite{jkim_327,marco_327}, we use the tetragonal notation [Figs. 1(a) and 1(b)] in presenting our RIXS results and define the momentum transfer $\textbf{Q}$ in reciprocal space as $\textbf{Q}=H\textbf{a}^\ast+K\textbf{b}^\ast+L\textbf{c}^\ast$, where $H$, $K$, and $L$ are Miller indices and ${\bf a}^\ast=\hat{{\bf a}}\frac{2\pi}{a}$, ${\bf b}^\ast=\hat{{\bf b}}\frac{2\pi}{b}$, ${\bf c}^\ast=\hat{{\bf c}}\frac{2\pi}{c}$ with $a\approx b\approx 3.9$ \AA, and $c \approx 20.9$ \AA\ for Sr$_3$Ir$_2$O$_7$. In this notation, the wave vector of the $c$-axis $G$-type AFM is $\textbf{q}=(\frac{1}{2}, \frac{1}{2}, 0)$ \cite{jkim_327,marco_327}.

We first describe the doping evolution of the magnetic order. Figure 1(a) shows the in-plane structure and the rotations of the IrO$_6$ octahedra in Sr$_3$Ir$_2$O$_7$. The exchange couplings on the tetragonal lattice and their naming conventions are described in Fig. 1(b). Figure 1(c) is a schematic magnetic and electronic phase diagram of (Sr$_{1-x}$La$_x$)$_3$Ir$_2$O$_7$, drawn according to previous works \cite{cao_la327,wilson_327}. The doping levels $x$ measured at $T = 30$ K are indicated by red circles. To characterize the doping dependent AFM, we have measured the magnetic Bragg peaks along $[H, \frac{1}{2}, 28]$ and $[\frac{1}{2}, \frac{1}{2}, L]$ for $x = 0, 0.02, 0.03, 0.05$, and $0.065$ using the elastic channel of the RIXS spectrometer \cite{SI}. The $L$ scan for $x = 0$ displays magnetic Bragg peaks from $L = 19$ to $28$ with an intensity modulation [Fig. 1(d)], which has a period controlled by the ratio between lattice parameter $c$ and bilayer distance ($d$) ($\frac{c}{d}\approx 5.1$) \cite{jwkim_327}. Upon electron doping, the 3D LAF persists for $x = 0.03$ but becomes short ranged for $x = 0.05$, as indicated by the broad peaks along both $H$ and $L$ in Figs. 1(e) to 1(f). The $L$ scan for the metallic $x = 0.05$ sample deserves special attention. It reveals a broad feature superimposed on a flat background, and is well fitted by a sum of the bilayer antiferromagnetic structural factor cos$^2(\frac{2\pi d}{c})$ \cite{jwkim_327} and a constant background [green solid curve in Fig. 1(f)]. The presence of this broad feature indicates that the magnetic correlation length along $c$ axis has decreased to a very small value comparable with the bilayer distance. This suggests that the $c$-axis magnetic correlations supporting the $G$-type AFM are on the verge of disappearing. The constant background can be attributed to a vanishing of the 3D SAF in a partial volume of the sample. This is in agreement with the percolative nature of the IMT, assuming that charge carriers are suppressing the magnetic order \cite{wilson_327}. For $x = 0.065$, the $L$ scan becomes featureless while the broader in-plane magnetic Bragg peak remains and persists at $290$K [Fig.1(e) and 1(f)]. This indicates that further doping suppresses the magnetic couplings that support the 3D SAF and drives the system into a robust 2D SAF state.

To understand the doping-dependent electronic interactions across the IMT and the transitions between 3D LAF and 2D SAF, we have presented the magnetic excitations and the spin-orbit excitons of (Sr$_{1-x}$La$_x$)$_3$Ir$_2$O$_7$ in Figs. 2 and 3. The in-plane momentum dependent RIXS for $x = 0, 0.02$ and $0.065$ are shown as color maps in Figs. 2(a)-(c). For $x = 0$, the dispersion, the large magnon gap and the spectral-weight distribution are consistent with a previous report measured at the same $L$ \cite{jkim_327}. In addition, our results reveal clear dispersive spin-orbit excitons exhibiting similar energy scale and dispersion with that in Sr$_2$IrO$_4$ \cite{dimensionality_mit, jkim_214, bjkim_14ncom, dingy_327, stefano_thesis}. As increasing $x$, the magnetic excitations are damped: they broaden in energy and decrease in intensity. This damping has been reported in other charge-carrier doped 2D correlated systems such as cuprates and iron pnictides \cite{tacon_11,dean_nmat_13, kejin_13, johnny_111, dai_rmp}. The spin-orbit excitons show weak doping dependence, indicating that the magnetic excitations are fluctuations of the robust $J_{\text{eff}} = \frac{1}{2}$ pseudospins \cite{SI}.

The strong doping dependence of the magnetic excitations are displayed in Figs. 2 and 3. The dispersions are symmetric about ($\frac{1}{4}, \frac{1}{4}$) and change less from $x = 0$ to $x = 0.03$ [Fig. 3(a)]. Across the IMT to $x = 0.05$, the dispersion becomes asymmetric and shows different gap size at ($\frac{1}{2}, \frac{1}{2}$) and ($0, 0$). A substantial softening occurs along ($\frac{1}{2}, \frac{1}{2}$)-($\frac{1}{4}, \frac{1}{4}$) while the band top at ($\frac{1}{2}, 0$) remains unchanged [Figs. 2 and 3(b)]. This anisotropic softening is followed by a further softening at ($\frac{1}{4}, \frac{1}{4}$) and, surprisingly, a sizable hardening at ($\frac{1}{2}, 0$) in $x = 0.065$ [Figs. 2 and 3]. Furthermore, the large magnon gap in $x \leq 0.05$ collapses dramatically in $x = 0.065$ [Figs. 2(f), 2(g) and 3(b)], where the magnetic excitations at ($\frac{1}{2}, \frac{1}{2}$) overlap with the elastic magnetic scattering [Fig. 2(f)], whereas a weak signal is observed at (0, 0) [Fig. 2(g)] \cite{SI}. A similar anisotropic softening between ($\frac{1}{4}, \frac{1}{4}$) and ($\frac{1}{2}, 0$) was observed in Sr$_{2-x}$La$_x$IrO$_4$ and attributed to the interaction between magnetic moments and emergent itinerant electrons having developed a Fermi surface at ($\frac{1}{4}, \frac{1}{4}$) \cite{La214_arpes,xliu_la214}. The same explanation can be applied here since La introduces itinerant electrons and Fermi pockets have been well developed at ($\frac{1}{4}, \frac{1}{4}$) in metallic (Sr$_{1-x}$La$_x$)$_3$Ir$_2$O$_7$ \cite{La327_arpes,junfeng_sr,junfeng_nmat}. Therefore the anisotropic softening is in line with that in Sr$_{2-x}$La$_x$IrO$_4$, indicative of a similar role of the emergent itinerant electrons.

The hardening of the magnetic excitations at ($\frac{1}{2}, 0$) and the collapse of the magnon gap occur between $x = 0.05$ and $0.065$ where the system evolves from 3D SAF to 2D SAF. Since the AFM along $c$-axis is absent in 2D SAF, we expect that the interlayer couplings will be greatly suppressed in $x = 0.065$.  With the suppression of the interlayer couplings, (Sr$_{1-x}$La$_x$)$_3$Ir$_2$O$_7$ will be much akin to the single-layer Sr$_{2-x}$La$_x$IrO$_4$ ($x \geq 0.04$), in which the magnetic excitations are gapless and have a larger band top at ($\frac{1}{2}, 0$) \cite{hlynur_la214}. Indeed, we find the 2D pseudospin-$\frac{1}{2}$ ($J$-$J_2$-$J_3$) model used for Sr$_2$IrO$_4$ captures the overall dispersion and generates similar fitted parameters ($J = 44, J_2 = -29, J_3 = 14.4$ meV) to Sr$_{2-x}$La$_x$IrO$_4$ ($x \geq 0.04$) [Fig. 3], indicating that the doped itinerant electrons drive the Sr$_3$Ir$_2$O$_7$ into a 2D magnetic system exhibiting strong antiferromagnetic pseudospin fluctuations \cite{hlynur_la214}. The strong magnetic excitations in $x = 0.065$ demonstrates that metallic (Sr$_{1-x}$La$_x$)$_3$Ir$_2$O$_7$ hosts strong electron correlations like its parent compound, and therefore provide a solid evidence for a correlated metallic picture \cite{wilson_327}.

\begin{figure}[htbp]
\includegraphics[width=8cm]{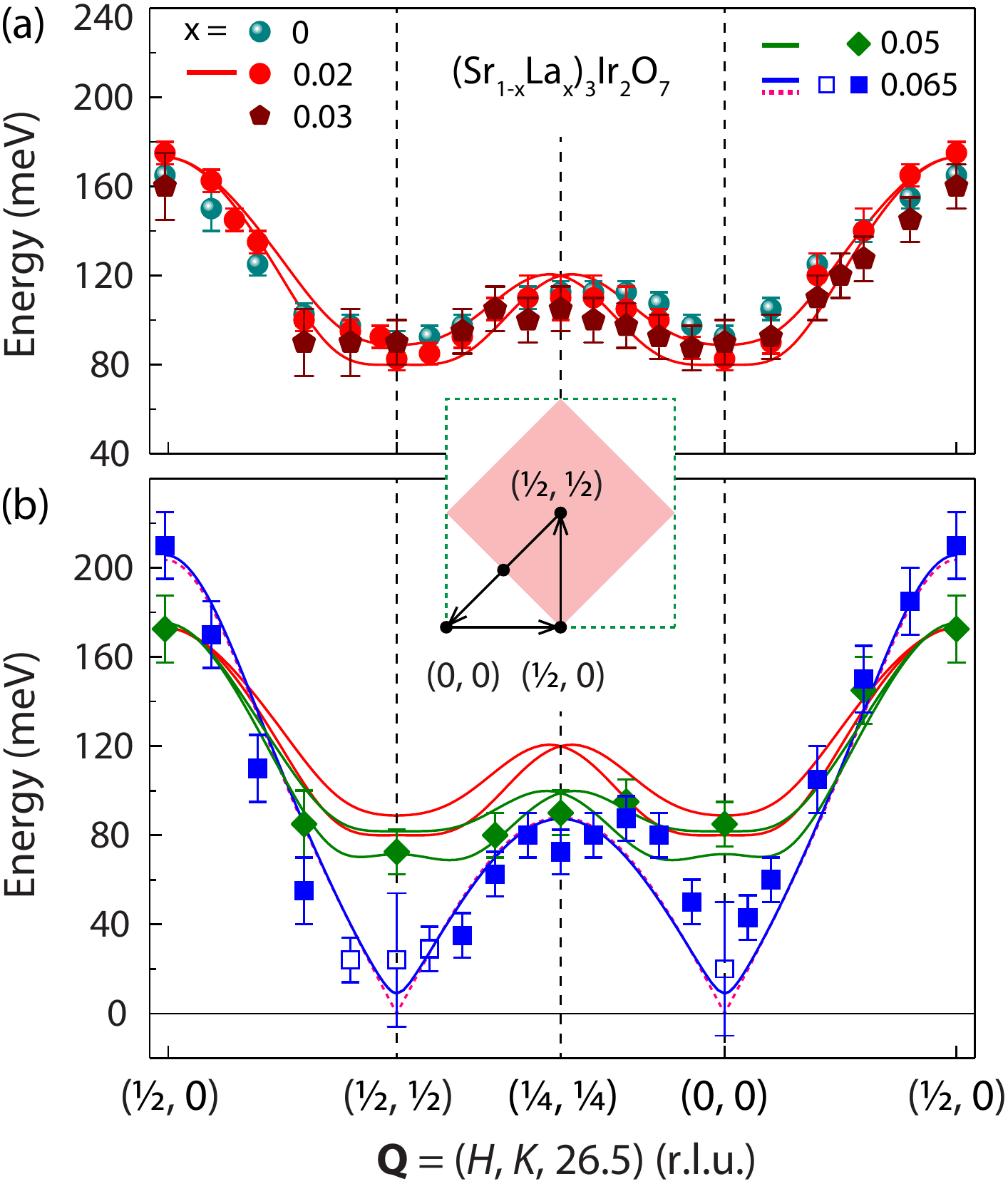}
\caption{(color online). Doping-dependent magnon dispersions for (Sr$_{1-x}$La$_x$)$_3$Ir$_2$O$_7$. To show the data clearly, the dispersions are split into two panels (a) $x = 0, 0.02$ and $0.03$ and (b) $x = 0.05$ and $0.065$. The dispersions for $x \leq 0.05$ and the blue solid squares of $x = 0.065$ are obtained by selecting the peak positions of the magnetic excitations. The blue open squares of $x = 0.065$ are extracted from fitting of the magnetic excitations \cite{SI}. The solid curves are fits to the dispersions for $x=0.02$, $0.05$ and $0.065$ using the bilayer model \cite{jkim_327}. The pink dashed curve is the fitting of the dispersion for $0.065$ using the $J$-$J_2$-$J_3$ model \cite{hlynur_la214}.}
\label{fig:3}
\end{figure}

To quantitatively understand the doping evolution of the magnetic couplings, we have fitted the dispersions using the bilayer model [Fig. 3] \cite{jkim_327, SI}, which consists of nearest neighbor ($H_{\text{0}}$) and long range interactions ($H_{\text{1}}$)
\begin{eqnarray}
{\it H_{\text{0}}} = \sum_{\langle i,j\rangle} \Bigl[ J_{ij}\vec S_i\cdot \vec S_{j}+\Gamma_{ij}S_{i}^{z}S_{j}^{z}
+{\vec D}_{ij}\cdot{\big (}\vec S_i\times\vec S_j{\big )} \Bigr]\,
\label{H0}
\end{eqnarray}

\noindent
\begin{eqnarray}
{\it H_{\text{1}}} = \sum_{\langle\langle i,j\rangle\rangle}J_2\vec{S_{i}}\vec{S_{j}}+\sum_{\langle\langle\langle i,j\rangle\rangle\rangle}J_3\vec{S_{i}}\vec{S_{j}}+ \sum_{\langle i,j\rangle}J_{2c}\vec{S_{i}}\vec{S}_{j+z}\;\,
\label{H1}
\end{eqnarray}

\noindent
where $\Gamma_{ij}$ is the anisotropic coupling including the bond-directional pseudodipolar interactions and $D_{ij}$ the Dzyaloshinsky-Moriya interaction arising from the staggered rotation of the IrO$_6$ octahedra [Fig. 1(a)]. $H_{\text{0}}$ contains the sum over both intralayer ($J_{ij}=J$, $\Gamma_{ij}=\Gamma$ and $D_{ij}=D$) and interlayer couplings ($J_c$, $\Gamma_c$ and $D_c$). The ${\langle i,j\rangle}, {\langle\langle i,j\rangle\rangle}$, and ${\langle\langle\langle i,j\rangle\rangle\rangle}$ in $H_{\text{1}}$ represents in-plane first, second and third nearest neighbors and $J_2$, $J_3$ and $J_{2c}$ the long range exchange couplings [Fig. 1(b)]. The $c$-axis couplings $J_c$ and $J_{2c}$ and $D_c$ are responsible for the bilayer splitting of the acoustic and optical branches, and the magnon gap arises from the anisotropic couplings $\Gamma$ and $\Gamma_c$ \cite{jkim_327}. Due to the presence of interlayer couplings, our fittings for $x \leq 0.05$ have been restricted to adjusting only $J$, $J_2$ and $J_3$ \cite{hlynur_la214, SI}. For $x = 0.065$, we fit all the parameters to account for the suppression of the interlayer couplings and the magnetic anisotropy. As shown in Fig. 3, our fitting successfully describes the anisotropic softening and the collapse of the magnon gap \cite{SI}. With increasing doping, the anisotropic softening is captured by the evolution of  $J$ and $J_2$ \cite{SI}, indicative of strong interplay between the in-plane nearest and next nearest couplings and the emergent itinerant electrons. The bilayer splitting of the two branches disappear in 2D SAF state since $J_c$, $J_{2c}$ and $D_c$ vanish. The collapse of the magnon gap corresponds to the suppression of the anisotropic couplings ($\Gamma$ and $\Gamma_c$) including the bond-directional pseudodipolar interactions \cite{SI}. With the suppression of the interlayer couplings, the bilayer model is similar to the $J-J_2-J_3$ model, as indicated by the fitting curves using these two models [Fig. 3(b)].

Although the models roughly describe the anisotropic softening and gap collapse, they fail to capture the asymmetry of the dispersions for metallic $x = 0.05$ and $0.065$ since these two models are intrinsically symmetric about ($\frac{1}{4}, \frac{1}{4}$). As shown in Fig. 3(b), the dispersion lies below the fitting along ($\frac{1}{2}, 0$)-($\frac{1}{2}, \frac{1}{2}$)-($\frac{1}{4}, \frac{1}{4}$) but above the fitting along ($\frac{1}{4}, \frac{1}{4}$)-($0, 0$)-($\frac{1}{2}, 0$). We attribute this to different damping rates between ($\frac{1}{2}, \frac{1}{2}$) and ($0, 0$) driven by the interactions between the pseudospins and the emergent itinerant electrons, which are not taken into account by the bilayer model developed based on the local-moment parent compound. Nonetheless, in phenomenological meaning, the effective fitting partially reflect the doping effects on the magnetic interactions (the dynamics of the pseudospins in itinerant context). On the other hand, we have also tentatively described the dispersions using the transverse mode of the quantum-dimer model reported in ref. \cite{marco_327}. The results have been discussed in the supplemental material \cite{SI}.

We now turn to the electron-doping effects on the spin-orbit excitons in (Sr$_{1-x}$La$_x$)$_3$Ir$_2$O$_7$. Besides demonstrating the robustness of the $j = \frac{1}{2}$ moments, the spin-orbit excitons undergo a sudden decrease in energy ($\Delta E_s \sim 70$ meV) at momentum close to ($\frac{1}{2}, 0$) between $x = 0.05$ and $0.065$. Since the spin-orbit excitons are electronic transitions from $j = \frac{3}{2}$ band to $j = \frac{1}{2}$ upper Hubbard band, this energy decrease suggests that these two bands are modified across the 3D-to-2D SAF transition \cite{SI}. This implies that the changes of magnetic order and magnetic couplings could induce band renormalization in this system \cite{SI}.

In summary, our RIXS measurements on (Sr$_{1-x}$La$_x$)$_3$Ir$_2$O$_7$ unveil an electron-doping driven evolution of the AFM from 3D LAF to 3D SAF and subsequent 2D SAF deep in the metallic state. Following this, we show that the magnons undergo an anisotropic damping with increasing doping, with the large magnon gap strongly suppressed in the 2D SAF metallic regime. This indicates that the emergent itinerant electrons suppress the AFM by weakening the magnetic couplings and drive the system into a 2D SAF correlated metallic state hosting strong AFM fluctuations of the $J_{\text{eff}} = \frac{1}{2}$ moments. Our results provide a solid experimental basis that will guide future theoretical works on the physics of doping the SOC-induced Mott insulators in the presence of strong interlayer couplings. In addition, the correlated metallic state hosting strong AFM pseudospin fluctuations in (Sr$_{1-x}$La$_x$)$_3$Ir$_2$O$_7$ could be a new platform for realizing novel quantum phases by applying internal or external perturbations.

\begin{acknowledgements}
We thank Matteo Rossi (ESRF) for helpful discussions. The work at PSI is supported by the Swiss National Science Foundation through its Sinergia network Mott Physics Beyond the Heisenberg Model(MPBH) and the NCCR-MARVEL. Xingye Lu acknowledges financial support from the European Community's Seventh Framework Program (FP7/2007-2013) under grant agreement NO. 290605 (COFUND: PSI-FELLOW). G. Cao acknowledges support by the US National Science Foundation via grants DMR-1265162 and DMR-1600057. The work used Sector 27 of the Advanced Photon Source, a U.S. Department of Energy (DOE) Office of Science User Facility operated for the DOE Office of Science by Argonne National Laboratory under Contract No. DE-AC02-06CH11357.
\end{acknowledgements}

{\it Note added}.--- Asymmetric dispersion of the magnetic excitations have recently also been observed by Hogan {\it et al}. in (Sr$_{1-x}$La$_x$)$_3$Ir$_2$O$_7$ \cite{hogan_16}, where the dispersion for $x = 0.07$ is much akin to our results on $x = 0.05$.

\end{document}